\documentclass[11pt,a4paper, onecolumn]{article}

\usepackage{url,colortbl}
\usepackage{xcolor}
\usepackage{bm,amsmath,amssymb}
\usepackage[]{graphicx}
\usepackage[margin=2cm]{geometry}
\usepackage[ruled,linesnumbered]{algorithm2e}
\usepackage{todonotes}
\graphicspath{{./Images/}}
\usepackage{authblk}
\usepackage{xr}
\usepackage{listings} 
\usepackage{dsfont}
\usepackage{appendix}
\usepackage[font=small]{caption}
\usepackage{setspace}
\usepackage{cite}
\usepackage{mathtools}
\newcommand{\corr}[1]{\textcolor{black}{#1}}

\begin{document}

\setlength{\parindent}{0pt}
\setlength{\parskip}{1em}

\doublespacing

\onecolumn

\renewcommand{\abstractname}{}
\title{The spontaneous emergence of leaders and followers in a \\ mathematical model of cranial neural crest cell migration}
\author[1]{Samuel W.S. Johnson\thanks{Corresponding author: samuel.johnson@chch.ox.ac.uk}}
\newcommand\CoAuthorMark{\footnotemark[\arabic{footnote}]}
\author[1]{Ruth E. Baker\thanks{These authors contributed equally.}}
\author[1]{Philip K. Maini\protect\CoAuthorMark}
\affil[1]{Wolfson Centre for Mathematical Biology, Mathematical Institute, University of Oxford, Oxford, United Kingdom}
\date{}
\maketitle
 
\linespread{1.5}\selectfont


\vspace{-36pt}

\begin{abstract}
    \noindent Many agent-based mathematical models of cranial neural crest cell (CNCC) migration impose a binary phenotypic \corr{partition} of cells into either leaders or followers. 
    In such models, the movement of leader cells at the front of collectives is guided by local chemoattractant gradients, while follower cells behind leaders move according to local cell--cell guidance cues. 
    Although such model formulations have yielded many insights into the mechanisms underpinning CNCC migration, they rely on fixed phenotypic traits that are difficult to reconcile with evidence of phenotypic plasticity \textit{in vivo}. 
    A later agent-based model of CNCC migration aimed to address this limitation by allowing cells to adaptively combine chemotactic and cell--cell guidance cues during migration. 
    In this model, cell behaviour adapts instantaneously in response to environmental cues, which precludes the identification of a persistent subset of cells as leader-like over biologically relevant timescales, as observed \textit{in vivo}. 
    \corr{Here, we build on previous leader--follower and adaptive phenotype models} to develop a polarity-based agent-based model of CNCC migration, in which all cells evolve according to identical rules, interact via a pairwise interaction potential, and carry polarity vectors that evolve according to a dynamical system driven by time-averaged exposure to chemoattractant gradients. 
    Numerical simulations of this model show that a leader--follower phenotypic \corr{partition} emerges spontaneously from the underlying collective dynamics of the model. Furthermore, the model reproduces behaviour that is consistent with experimental observations of CNCC migration in the chick embryo. 
    Thus, we provide an experimentally consistent, mechanistically-grounded mathematical model that captures the emergence of leader and follower cell phenotypes without their imposition \textit{a priori}.

\end{abstract}

\newpage

\linespread{2}\selectfont

\section{Introduction}

The collective migration of cranial neural crest cells (CNCCs) is a fundamental process in the embryonic development of vertebrates \cite{kulesa2010cranial}. 
During embryogenesis, CNCCs migrate collectively away from the embryonic hindbrain towards migratory targets in branchial arches 1--4. 
When CNCCs colonise the branchial arches, they subsequently differentiate to facilitate the formation of the craniofacial structure in the embryo \cite{cordero2011cranial, santagati2003cranial}.  
Aberrant CNCC migration is associated with a range of birth defects, termed \textit{neurocristopathies}, such as cleft palate \cite{ito2003conditional} and Treacher Collins syndrome \cite{dixon2007treacher}. 
Insights into the mechanisms underpinning CNCC migration may, therefore, provide opportunities to expedite the development of medical interventions that aim to correct disruptions to migration and the resultant craniofacial birth defects.

Chick CNCCs migrate up self-induced gradients of the chemoattractant VEGF, which guide their migration towards the branchial arches \cite{mclennan2010vascular}. 
In addition to VEGF signalling, CNCC migration is influenced by extracellular matrix (ECM) components such as laminin and fibronectin \cite{newgreen1980fibronectin, duband1987distribution}, and by local cell--cell interactions that coordinate collective motion \cite{taneyhill2008adhere}. 
The relative importance of these guidance cues varies between cells within migrating streams, as revealed by genetic analyses in the chick embryo;
cells at the leading edge of streams that are exposed to steep VEGF gradients upregulate genes associated with VEGF signalling and ECM attachment \cite{mclennan2015vegf, mclennan2012multiscale}. Conversely, cells in trailing regions, experiencing shallower gradients, upregulate genes associated with cell--cell adhesion and contact guidance \cite{mclennan2012multiscale}. 
Despite these well-characterised molecular and behavioural heterogeneities, it remains unclear to what extent CNCC migratory behaviours exist along a continuous phenotypic spectrum or constitute a discrete leader--follower dichotomy, and how such phenotypic differences may arise during migration.

In the study of CNCC migration, mathematical modelling offers a valuable addition to experiments, providing a means of both rapidly testing existing biological hypotheses and generating novel hypotheses that may later be tested with experiments. 
Many agent-based models (ABMs) of CNCC migration approximate the genotypic variation observed across CNCC streams \textit{in vivo} by imposing a binary phenotypic \corr{partition} between leaders, which move according to gradients in VEGF sensed by filopodia, and followers, which move according to local guidance cues from adjacent cells \cite{mclennan2012multiscale, mclennan2015vegf, mclennan2023colec12, mclennan2020neural, mclennan2017dan, johnson2025mathematical}. 
Such models can recapitulate many CNCC behaviours observed \textit{in vivo}, and have yielded insights into both the mechanisms through which CNCCs migrate and the environmental factors influencing their migration. 

However, the binary phenotypic \corr{partition} between leader and follower cells present in many ABMs of CNCC migration precludes a more in-depth investigation into phenotypic plasticity and variation across CNCC streams that likely alters the migratory behaviour of collectives. 
An \corr{alternative} ABM of CNCC migration aimed to address these limitations by allowing cells to respond to both cell--cell guidance cues and chemoattractant gradients simultaneously \cite{schumacher2019neural}. 
\corr{In this model, the movement of cells is determined by a combination of leader-like chemical cues from VEGF gradients and follower-like cell--cell guidance cues, with their weighting determined by the instantaneous strength of the VEGF gradient detected by cells.} 
In relaxing the binary phenotypic \corr{partition} considered in other models, this model allows for a more in-depth exploration of adaptive guidance cues in CNCC migration, and, under certain conditions, produces streams consistent with those observed \textit{in vivo}. 
\corr{However, the behaviour of each cell in this framework} depends on the instantaneous cues detected in its local environment, such that the relative contribution of chemotaxis and cell--cell guidance in movement may rapidly fluctuate between leader-like and follower-like throughout migration. 

In this study, we investigate the mechanisms underlying the emergence of distinct phenotypes during CNCC migration and the extent to which these emergent phenotypes conform to a leader--follower \corr{dichotomy}.
We formulate a novel \textit{spring--polarisation} ABM of CNCC migration, wherein cell motion is determined by a pairwise interaction potential and a cell--specific polarity vector that evolves in orientation and magnitude according to time-averaged gradients in VEGF \corr{concentration} surrounding cells. 
Numerical simulations of this model show that a leader--follower phenotypic \corr{partition} emerges spontaneously from the underlying dynamics of the model, according to the magnitude of the chemical polarisation vector throughout migration. 
Additionally, the model reproduces phenomena consistent with prior experiments conducted in the chick embryo, including the resumption of migration upon the ablation of leading cells \cite{richardson2016leader} and the transplantation of follower cells ahead of the leading edge of streams \cite{mclennan2012multiscale}. 
Thus, our model represents a significant development on prior ABMs of CNCC migration \cite{mclennan2012multiscale, mclennan2015vegf, mclennan2023colec12, mclennan2020neural, mclennan2017dan, johnson2025mathematical}, in that stable leader and follower phenotypes are shown to emerge spontaneously from the underlying model dynamics without their imposition \textit{a priori}, \corr{with streams remaining capable of reorganising in response to environmental perturbations}.

Biologically, our results show that leader--follower organisation in the cranial neural crest can emerge purely from collective dynamics within CNCC streams. 
Persistent polarisation of CNCCs in strong VEGF gradients at the leading edge gives rise to stable leader-like behaviour. CNCCs in trailing positions, exposed to comparatively weaker VEGF gradients, adopt follower-like phenotypes in which cells remain largely unpolarised by \corr{shallow} gradients in VEGF and instead depend strongly on local cues from adjacent cells for their movement.
These findings support the notion of an approximate leader--follower phenotypic division in the cranial neural crest, but suggest that this may emerge during migration, rather than being predetermined before migration begins.

\section{Results}

We formulate the spring--polarisation ABM of CNCC migration, before conducting numerical simulations to study the model behaviour as a function of various biological parameters and experimental perturbations. 

\subsection{Spring--polarisation ABM of CNCC migration}\label{sec:model}

\begin{figure}[ht!]
    \centering
    \includegraphics[width=\linewidth]{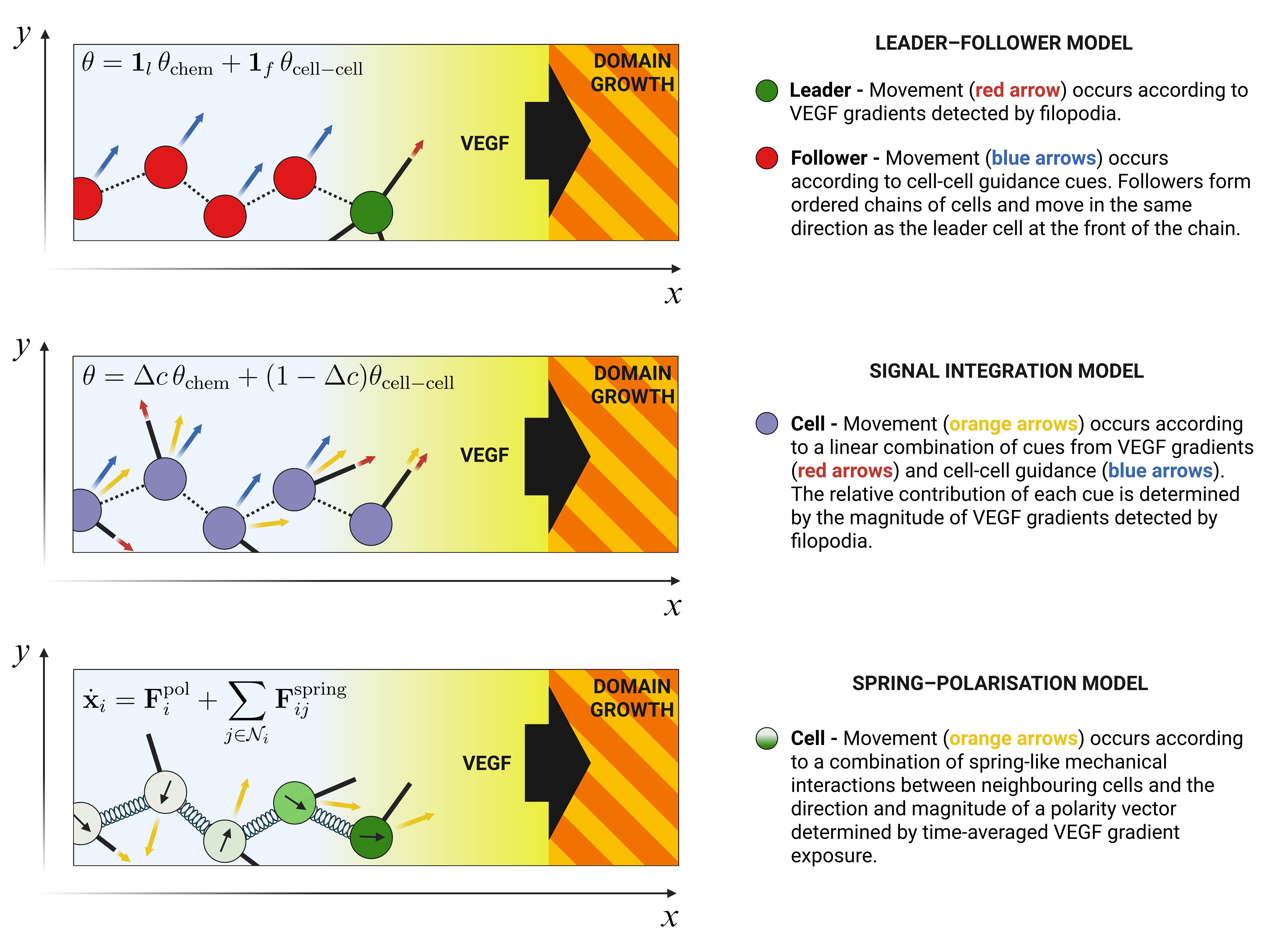}
    \caption{Schematics of previous leader–follower \cite{mclennan2012multiscale, mclennan2023colec12, mclennan2017dan, mclennan2015vegf, johnson2025mathematical} and signal integration \cite{schumacher2019neural} models, together with the spring–polarisation model of CNCC migration formulated here. In leader--follower models, a fixed subset of leader cells at the front of streams move according to VEGF gradients detected in the cranial micro-environment. Follower cells, in weaker gradients of VEGF, move according to cell--cell guidance cues, forming connected chains of follower cells led by a leader cell at the front. In the signal integration model, all cells migrate according to a combination of chemical and cell--cell guidance cues. The movement of each cell is determined by a linear combination of these two mechanisms of movement, with the relative contribution of each mechanism governed by the magnitude of the chemical gradient detected at the time of movement. In the spring--polarisation model, all cells interact with their neighbours via spring-like mechanical interactions and carry a cell-specific polarity vector that evolves in time according to VEGF gradient sensing. Movement occurs according to a combination of forces from spring-like interactions and a force determined by the orientation and magnitude of the polarity vector.}
    \label{fig:Figure1}
\end{figure}

We formulate a stochastic ABM representing the migration of chick CNCCs over a period of $24$h. In the model, CNCCs are modelled as point particles whose movement is governed by spring-like pairwise mechanical interactions with their neighbours and the magnitude and orientation of a cell-specific polarity vector that evolves according to time-averaged exposure to VEGF gradients. 
Here, the cell-specific polarity vector represents the response of CNCCs to timed exposure to VEGF, which has been shown to induce rapid phenotypic transitions \textit{in vitro} \cite{mclennan2015vegf}. 
In this model formulation, we avoid the \textit{a priori} prescription of cell phenotypes used in prior ABMs of CNCC migration \cite{mclennan2012multiscale, mclennan2023colec12, mclennan2017dan, mclennan2015vegf, johnson2025mathematical}, in that all cells are governed by the same set of movement rules. 
However, we allow for the emergence of distinct phenotypes through differences in the time evolution of VEGF-induced polarity in each cell, which determines the relative contributions of chemotaxis and cell--cell interactions towards movement. 
As in the prior signal integration ABM of CNCC migration \cite{schumacher2019neural}, cells move according to a combination of cues from cell--cell interactions and VEGF gradients, though in the model considered here, cell movement depends on the results of environmental sampling over a finite time interval, rather than on instantaneous cues detected at the time of movement. 
In doing so, we introduce a notion of memory into CNCC movement, which, in turn, allows individual CNCCs to integrate fluctuating local cues into migratory behaviours that persist over time. 
This memory-dependent movement mechanism enables transient heterogeneities in VEGF exposure or cell density to generate sustained collective migration patterns without the explicit assignment of leader or follower phenotypes. 
Consequently, the model \corr{demonstrates how} coordinated CNCC migration and phenotypic variation \corr{can be viewed as} emergent properties of the underlying chemical and mechanical movement mechanisms.
A visual summary of previous leader--follower and signal integration models, alongside a schematic of the spring--polarisation model formulated here, is shown in Figure \ref{fig:Figure1}. 

\subsubsection{Migratory domain}\label{sec:modelDomain}
We model CNCC migration on a two-dimensional domain, $(x,y)\in L_{x}(t)\times L_{y}$, that grows in length over the $24$h period of migration (Figure \ref{fig:sweep}A). 
The domain represents the  region of the embryo between rhombomere 4 (r4) and branchial arch 2 (ba2), the r4--ba2 pathway, which is a region of the vertebrate embryo through which CNCC migration occurs \cite{mclennan2012multiscale}. 
The domain is approximated as a rectangle of a fixed width, $L_y = 120\mu$m, and is of a length, $L_{x}(t)$, that increases over a period of $24$h, from approximately $300\mu$m at the start of migration to $1100\mu$m when migration ends. 
As in prior models of CNCC migration \cite{mclennan2012multiscale, mclennan2023colec12, mclennan2017dan, mclennan2015vegf, johnson2025mathematical}, the growth of the domain is approximated as spatially-uniform. 
The length of the domain, $L_{x}(t)$, is derived from \textit{in vivo} data collected in the chick embryo \cite{mclennan2012multiscale}. Further details of the method through which $L_{x}(t)$ was obtained from these data are available in Appendix~A.1.

\subsubsection{Chemoattractant dynamics}\label{sec:chemDynamics}
In the two-dimensional domain through which migration occurs, we represent VEGF as a diffusible factor, $c(x,y,t)$, that is initially present everywhere in the domain. When migration begins at $t=0$h, the concentration of VEGF is spatially uniform ($c(x,y,0) = 1$ in non-dimensional units). The concentration of VEGF at later times is governed by a partial differential equation (PDE) that depends on the position of CNCCs in the domain and that has been used in prior models of CNCC migration \cite{mclennan2012multiscale, mclennan2023colec12, mclennan2017dan, mclennan2015vegf, johnson2025mathematical, schumacher2019neural}. The PDE includes terms for diffusion, production, degradation by CNCCs, and dilution due to domain growth. After re-scaling to a domain of unit length ($(x, y) \in (0, 1) \times (0, L_{y})$) for numerical convenience, we obtain

\small
\begin{equation}
{\fontsize{9}{11}\selectfont
    \label{chemicalRescaledEquation}
    \frac{\partial c}{\partial t} =
    \underbrace{\vphantom{c\sum_{i=1}^{N(t)} \frac{\lambda_{i}}{2\pi R^{2}}
      \exp\!\left[-\frac{L_{x}(t)^{2}(x-x_{i})^{2}+(y-y_{i})^{2}}{2R^{2}}\right]}
      \smash{D\left( \frac{1}{L_{x}(t)^{2}}\frac{\partial^{2} c}{\partial x^{2}}
      + \frac{\partial^{2} c}{\partial y^{2}}\right)}}_{\text{(I)}}
    -
    \underbrace{c\sum\nolimits_{i=1}^{N(t)} \frac{\lambda}{2\pi R^{2}}
      \exp\!\left[-\frac{L_{x}(t)^{2}(x-x_{i})^{2}+(y-y_{i})^{2}}{2R^{2}}\right]}_{\text{(II)}}
    +
    \underbrace{\vphantom{c\sum_{i=1}^{N(t)} \frac{\lambda_{i}}{2\pi R^{2}}
      \exp\!\left[-\frac{L_{x}(t)^{2}(x-x_{i})^{2}+(y-y_{i})^{2}}{2R^{2}}\right]}
      \smash{\kappa\, c(1 - c)}}_{\text{(III)}}
    -
    \underbrace{\vphantom{c\sum_{i=1}^{N(t)} \frac{\lambda_{i}}{2\pi R^{2}}
      \exp\!\left[-\frac{L_{x}(t)^{2}(x-x_{i})^{2}+(y-y_{i})^{2}}{2R^{2}}\right]}
      \smash{\frac{c}{L_{x}}\,\dot{L}_{x}(t)}}_{\text{(IV)}}. 
      }
\end{equation}

\normalsize

In Equation~\eqref{chemicalRescaledEquation}, term (I) describes diffusion with diffusion \corr{coefficient} $D$. Term (II) captures VEGF degradation mediated by CNCCs. This degradation occurs at a maximum rate $\lambda$ within a characteristic radius $R$ of a cell centred at $(x_i,y_i)$, for each of the $N(t)$ CNCCs in the population. Term (III) specifies spatially uniform VEGF production at a rate $\kappa$, and term (IV) accounts for dilution arising from domain growth \corr{(where the dot denotes differentiation with respect to time)}. Zero-Dirichlet boundary conditions are applied at $x = 0$ and $x = 1$ to prevent artificial polarisation towards the boundaries as cells enter the domain, while periodic boundary conditions are imposed at $y = 0$ and $y = L_{y}$ to represent the presence of adjacent tissue segments, into which VEGF may diffuse. The non-rescaled form of Equation~\eqref{chemicalRescaledEquation} is given in Appendix~A.2, and parameter values are based on previous ABMs of CNCC migration, as detailed in Appendix~B.

\subsubsection{Cell sensing, polarity, and heading dynamics}\label{sec:polarity}

Each cell, indexed by $i$, is represented as an autonomous agent that samples its local environment 
through a fixed number of filopodia, which are used to infer the direction of the steepest local 
positive VEGF gradient.

In the chick embryo, CNCCs do not begin large-scale migration immediately upon specification. Instead, delamination from r4 and the onset of migration occur only after a short, well-characterised lag of several hours \cite{theveneau2012neural}. Accordingly, no cells are present in the domain between $0 \le t < 6$h. At $t=6$h, a small cohort of cells is initialised along the domain boundary at $x=0$, with evenly spaced $y$-coordinates spanning $y \in [0, L_y]$. Thereafter, new cells are proposed for insertion at $x=0$ in intervals of $1/60$h, each with a randomly assigned $y$-coordinate. A proposed insertion is accepted only if the inserted cell lies at least two cell interaction lengths ($15\mu$m) from all other cells, where the interaction 
length defines the effective range of the short-range repulsive interaction potential between cells (Section \ref{sec:motion}). The continual addition of cells during migration represents the ongoing delamination of CNCCs from the embryonic hindbrain during development~\cite{kulesa2010cranial}.

At each time step, every cell samples $n_{\rm f}=5$ orientations, $\{\phi_k\}_{k=1}^{n_{\rm f}}$, within a circular region of a radius equal to the filopodial length, $\ell_{\rm f}=27.5\mu$m.
These values of $n_{\rm f}$ and $\ell_{\rm f}$ are \corr{comparable} to previous models of CNCC migration~\cite{mclennan2012multiscale, mclennan2015vegf, schumacher2019neural, mclennan2017dan, johnson2025mathematical, mclennan2020neural, mclennan2023colec12} and are informed by \textit{in vivo} and \textit{in vitro} measurements from the chick embryo~\cite{mclennan2012multiscale}.
Although the number of sampled orientations, $n_{\rm f}=5$, is lower than the number of filopodia typically extended by CNCCs, this is offset by the fact that new orientations are sampled at every time step ($1/60$~h) in the model, whereas \textit{in vivo}, filopodia persist over longer time intervals \cite{teddy2004vivo}.

For each orientation $\phi_k$ at each time step, each cell computes the mean VEGF concentration along the corresponding filopodium

\begin{equation}
c_{\mathrm{f}}(\phi_k) = \frac{1}{\ell_{\rm{f}}}\int_0^{\ell_{\rm{f}}} 
c\!\left(\mathbf{x}_i + r(\cos\phi_k,\sin\phi_k)\right)\,{\rm d}r,
\end{equation}

and compares this to the concentration of VEGF at its current position, $c(\mathbf{x}_i)$.  
The sampled direction that maximises $c_{\mathrm{f}}(\phi_k)$, $\phi_{\max} = \arg\max_{\phi_k \in \{\phi_1,\dots,\phi_{n_{\rm f}}\}} c_{\mathrm{f}}(\phi_k)$, defines an instantaneous chemotactic direction 
\(\hat{\mathbf{u}}_{\rm chemo} = (\cos\phi_{\max},\sin\phi_{\max})\).

A relative cue, $\sigma$, and detection threshold, $\upsilon$, based on prior ABMs (e.g.~\cite{mclennan2012multiscale}), are then computed as
\begin{equation}
\sigma = 
\frac{c_{\rm f}(\phi_{\rm{max}})-c(\mathbf{x}_i)}{c(\mathbf{x}_i)},
\qquad 
\upsilon = 
\xi\,\sqrt{\frac{c_0}{c(\mathbf{x}_i)}},
\end{equation}
where $c_0$ is the initial concentration of VEGF in the domain, and $\xi>0$ is a dimensionless parameter that controls the sensitivity threshold for gradient detection.  
The chemotactic activation variable, $\chi \in [0,1]$, quantifies the strength of a cell’s response to detected chemical cues.  
It is computed from the relative concentration signal, $\sigma$, and the sensitivity threshold, $\upsilon$, such that chemotactic responses are inactive when the chemotactic signal is below a threshold value and increase linearly when the signal is above it, before saturating at unity
\begin{equation}
\chi =
\begin{cases}
0, & \sigma \le \upsilon,\\[4pt]
\min\!\left(1,\;\dfrac{\sigma-\upsilon}{\gamma_{\mathrm{sat}}\upsilon}\right), & \sigma>\upsilon.
\end{cases}
\end{equation}

The parameter $\gamma_{\mathrm{sat}}$, therefore, determines how rapidly chemotactic responses saturate at a maximum once the threshold for gradient detection is exceeded. In simulations, we fix $\gamma_{\mathrm{sat}}=0.5$, meaning a maximum chemotactic response is achieved when detected gradients are at least 1.5 times the magnitude of the threshold for gradient detection. This is an arbitrary parameter choice used to reduce the dimensionality of the free model parameter space. However, all trends discussed in later sections are robust to variations in this parameter. 

Each cell stores a two-dimensional polarity memory vector, $\mathbf{m}_i$, that evolves according to an ordinary differential equation (ODE) with a characteristic relaxation time, $\tau_{\rm sense}>0$,

\begin{equation}
\label{mEq}
\dot{\mathbf{m}}_i(t)
= \frac{1}{\tau_{\mathrm{sense}}}
\bigl(\chi\,\hat{\mathbf{u}}_{\mathrm{chemo}}
- \mathbf{m}_i(t)\bigr), 
\end{equation}

where $\hat{\mathbf{u}}_{\mathrm{chemo}}$ is the direction vector of the highest measured VEGF gradient. Similar representations of cell polarity dynamics have been implemented in other mathematical models of collective cell chemotaxis (e.g. \cite{camley2016emergent}). In simulations, we set $\tau_{\mathrm{sense}} = 0.1$h, corresponding to a characteristic timescale of approximately six minutes for polarity reorientation.
This parameter choice reflects the rapid VEGF-dependent transcriptional and behavioural responses observed \textit{in vitro}, where gene-expression changes were detected within less than ten minutes of VEGF withdrawal or re-exposure~\cite{mclennan2015vegf}.

At each time step, the magnitude of the memory vector, $\|\mathbf{m}_i\|$, determines the magnitude of active forces generated by chemotactic gradients, while the memory angle, 
\(\theta_{\rm mem}=\arg(\mathbf{m}_i)\), sets the preferred heading for chemotactic movement.  
The instantaneous cell orientation, \(\theta_i\), relaxes towards  \(\theta_{\rm{mem}}\) with an angular rate \(k_{\rm chemo}\)

\begin{equation}
\dot{\theta}_i(t)
= k_{\mathrm{chemo}}\sin\bigl(\theta_{\mathrm{mem}}(t) -\theta_i(t)\bigr),
\qquad
\theta_i(t) \in \mathbb{R}/2\pi\mathbb{Z},
\end{equation}

which is an angular relaxation model that has previously been used in models of cell migration and polarity \cite{sliusarenko2007aggregation, zhang2018agent}.  In simulations, we set $k_{\mathrm{chemo}} = 10\mathrm{h}^{-1}$, corresponding to a characteristic relaxation timescale of approximately six minutes. This value once again reflects the rapid transcriptional responses observed in CNCCs exposed to ectopic sources of VEGF \textit{in vitro} \cite{mclennan2015vegf} and is also consistent with ABM simulations that predict maximal migratory efficiency in collectives when phenotype switching occurs on the order of minutes \cite{mclennan2015vegf}. Further details on the implementation of chemical polarisation in numerical model simulations are given in Appendix~A.3. 

\subsubsection{Over-damped active mechanics}
\label{sec:motion}

Cells are represented as point particles located at positions $\mathbf{x}_i=(x_i,y_i)$ within the two-dimensional migratory domain, $L_x(t) \times L_y$.  
Movement occurs in the over-damped limit, where inertial effects are negligible.  
The instantaneous velocity of each cell satisfies  

\begin{equation}
\dot{\mathbf{x}}_i
=
\mathbf{F}^{\mathrm{act}}_i
+ \sum_{\substack{j \in \mathcal{N}_i \\ j \ne i}}
\left(
\mathbf{F}^{\mathrm{spring}}_{ij}
+ \mathbf{F}^{\mathrm{rep}}_{ij}
\right),
\end{equation}

where $\mathbf{F}^{\mathrm{act}}_i$ is the polarity-driven active force, $\mathcal{N}_i$ is the neighbourhood of cell $i$, $\mathbf{F}^{\mathrm{spring}}_{ij}$ represents an effective adhesive interaction between neighbouring cells (for example, Cadherin-mediated adhesion between CNCCs \cite{mclennan2012multiscale, morrison2017single}), and $\mathbf{F}^{\mathrm{rep}}_{ij}$ is a short-range interaction force that imposes an effective exclusion zone around cells (and thus represents the finite size of cells \textit{in vivo}). In numerical simulations, stochasticity in movement is introduced through an additive positional perturbation applied at each discrete time step of the overdamped cell movement (Appendix B.1.4).

The active force on a cell due to chemotaxis is proportional to the magnitude of the cell-specific polarity vector, $||\mathbf{m}_{i}||$, and directed along the current heading
\begin{equation}
    \mathbf{F}^{\rm act}_i = f_0\, ||\mathbf{m}_{i}||\,(\cos\theta_i,\sin\theta_i),
\end{equation}

where $f_0$ is a parameter that defines the characteristic chemotactic force strength per cell.

Adhesion between neighbouring cells is modelled using dynamic Hookean springs of stiffness $k_{\rm spring}$, that represent transient adhesive bonds between cells. When the distance between the centres of two cells falls below a linking radius corresponding to the maximum distance of cell--cell interactions, $\ell_{\rm f}^{\max}$, an adhesive bond is formed and a spring is introduced between the cells. The rest length of the spring, $L_0$, is set to the inter-cell distance at the time of bond formation.

If \(r_{ij}\) denotes the distance between cells \(i\) and \(j\), and \(\hat{\mathbf{n}}_{ij}\) is the unit vector from cell \(i\) to cell \(j\), the adhesive force acting on cell \(i\) due to cell \(j\) is given by
\begin{equation}
\mathbf{F}^{\mathrm{spring}}_{ij}
=
k_{\mathrm{spring}}\,(r_{ij}-L_0)\,\hat{\mathbf{n}}_{ij}.
\end{equation}
Bonds are broken when the inter-cell distance exceeds a prescribed breaking radius, taken to be slightly larger than the linking radius (Appendix~B.1.5). This implementation provides a coarse-grained approximation of cell--cell adhesive interactions arising from effects such as cadherin-mediated contacts \cite{taneyhill2008adhere} and cell--cell interface deformation \cite{maitre2011role}.

Volume exclusion is represented by a soft-core repulsive force when cells overlap 
\begin{equation}
\mathbf{F}^{\rm rep}_{ij}
=
\begin{cases}
-\,k_{\rm rep}\,(2R-r_{ij})\,\hat{\mathbf{n}}_{ij},
& r_{ij}<2R,\\[4pt]
\mathbf{0}, & r_{ij}\ge 2R.
\end{cases}
\end{equation}
Zero-flux boundary conditions are imposed on cells, such that they remain within the domain throughout migration, thus representing the presence of inhibitory signals adjacent to the r4--ba2 pathway \textit{in vivo} \cite{mclennan2023colec12, johnson2025mathematical}.

Here, it is important to highlight that all mechanisms of environmental and intercellular interactions in this model are isotropic, in that there is no prescribed spatial bias in cell sensing or movement. This represents a relaxation of assumptions made in previous ABMs of CNCC migration \cite{mclennan2012multiscale, mclennan2023colec12, mclennan2017dan, mclennan2015vegf, johnson2025mathematical, schumacher2019neural}, where movement cues must be taken from cells ``ahead'' in follower cell chains (i.e. those closer to leader cells), but not from those ``behind'' (those further away from leader cells), in order for directional collective migration to occur. 
With the model formulated here, we show in later sections that the biased migration of CNCC collectives is a phenomenon that can occur solely due to an emergent asymmetry in VEGF profiles, with no prescribed orientational asymmetry in cell--cell interactions. In the model, asymmetric cell-induced VEGF gradients induce a non-zero polarity in the positive $x$-direction for cells at the leading edge of streams, which, through spring-like mechanical couplings, induces a subsequent directional force on trailing cells. This effect results in migration as a collective without an in-built spatial bias in cellular responses to environmental cues.
This, in turn, suggests that polarisation and directional movement are qualities that may propagate along streams of CNCCs from cells at the leading edge of collectives through local intercellular interactions that transmit alignment cues during migration.

\subsection{Collective CNCC migration requires balanced intercellular adhesion, active polarity-driven forces, and CNCC-induced VEGF degradation}

In model simulations, the three parameters that most strongly influence collective CNCC migration are $f_0$, which scales the magnitude of active forces generated by VEGF-induced polarisation, 
$k_{\mathrm{spring}}$, the Hookean spring constant governing intercellular attraction, and $\lambda$, the rate at which VEGF is degraded by cells. 

\begin{figure}
    \centering
    \includegraphics[width=\linewidth]{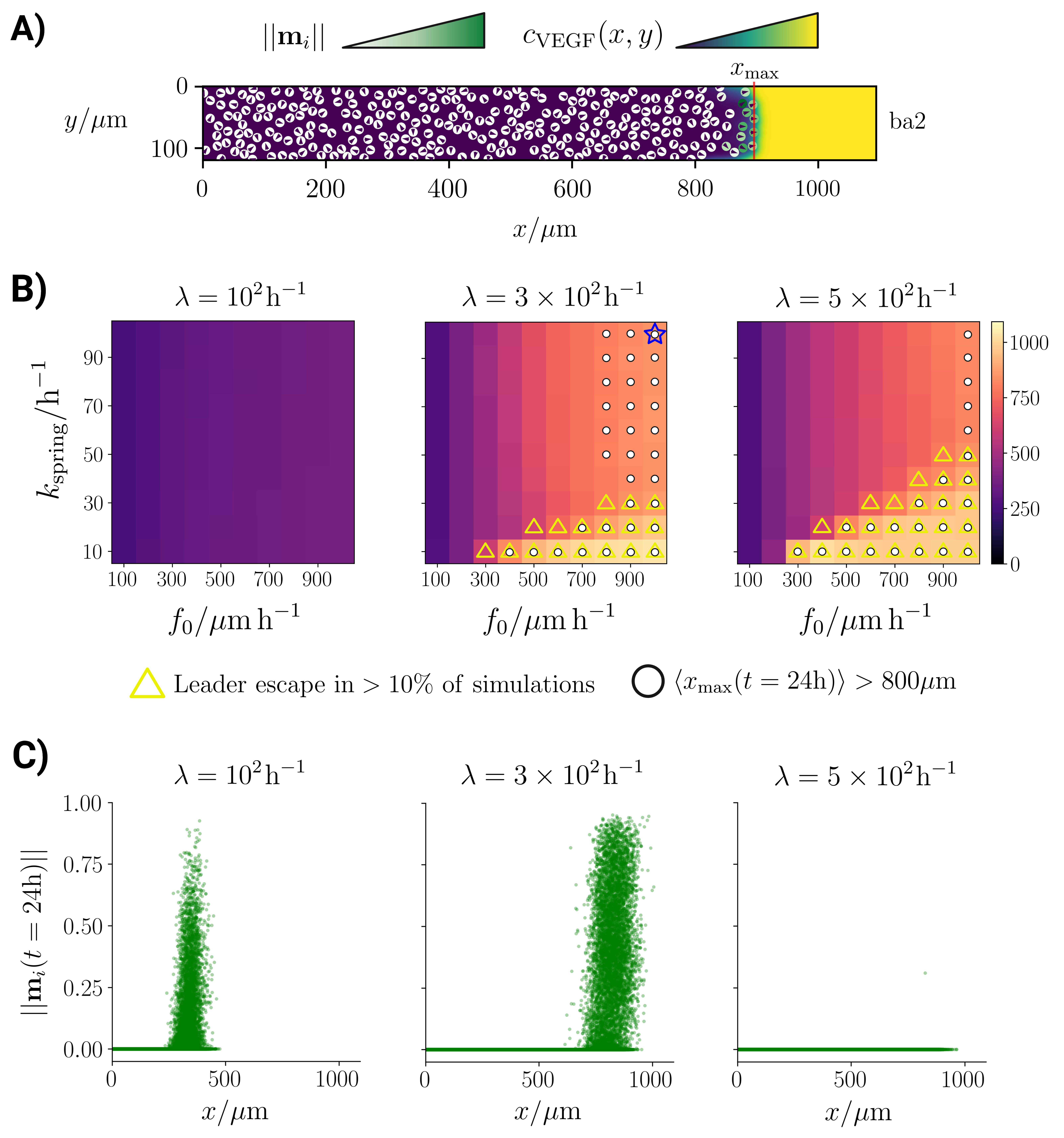}
    \caption{A) Example model simulation ($t=24\rm{h}$) for the parameter set $\{f_0=1000\mu\rm{m\,h^{-1}}, k_{\mathrm{spring}}=100\rm{h^{-1}}, \lambda=300\rm{h^{-1}}\}$. Non-zero polarisation ($||\mathbf{m}_{i}|| > 0$) at the leading edge of the stream is maintained due to strong local VEGF gradients, driving collective migration towards ba2 over a biologically relevant distance. The statistic $x_{\rm{max}}$ denotes the maximum $x$-coordinate across all cells in the stream when migration ends. B) $\langle x_{\rm{max}}\rangle$ as a function of $\lambda$, $f_0$, and $k_{\rm{spring}}$. Yellow triangles denote regions of parameter space in which $>10\%$ of model simulations result in leader escape, where at least one cell lies more than three cell radii ahead of the bulk when migration ends. White circles denote regions of parameter space in which $\langle x_{\rm{max}}\rangle$ exceeds $800\mu$m. The blue star denotes the representative parameter set that is used to analyse model behaviour in Sections \ref{emergence} and \ref{experiments}. C) Scatter plots of $||\textbf{m}_{i}||$ as a function of $x$ at $t=24$h as for small, intermediate, and large values of $\lambda$ for the representative parameter set $\{f_0=1000\mu\rm{m\,h^{-1}}, k_{\mathrm{spring}}=100\rm{h^{-1}}\}$. The results for each parameter combination are averaged over $500$ simulations.}
    \label{fig:sweep}
\end{figure}

Numerical simulations of the spring--polarisation model demonstrate that moderate values of $\lambda$ are required for sustained collective CNCC migration over biologically relevant distances (Figure \ref{fig:sweep}B, C). If $\lambda$ is too small, cells at the leading edge of streams are exposed only to shallow gradients in VEGF, and thus remain weakly-polarised throughout migration, generating active forces that are insufficient to drive the migration of collectives over biologically relevant distances (Supplementary Video S1). Conversely, if $\lambda$ is too large, rapid VEGF degradation erodes the local chemotactic gradients that are required to maintain non-zero polarity at the leading edge of streams, resulting in a transient period of high polarity before a loss of polarisation at the leading edge and a subsequent cessation of active motility (Supplementary Video S2). Only when $\lambda$ takes intermediate values, are stable VEGF gradients preserved that sustain sufficient polarity in cells at the leading edge of streams and, thus, continued active migration towards ba2. 

\begin{figure}
    \centering
    \includegraphics[width=\linewidth]{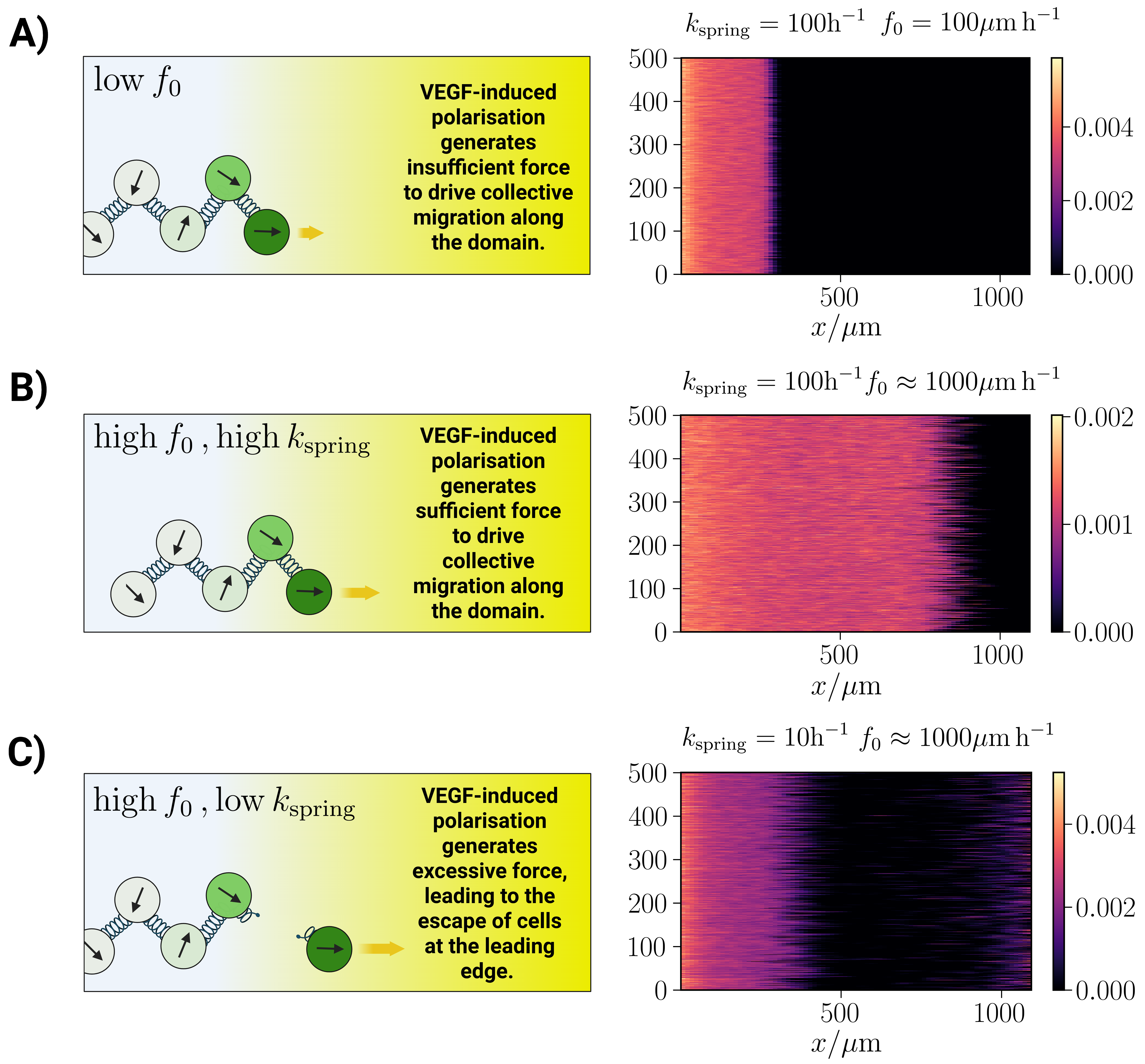}
    \caption{Schematics of the relationship between $f_0$ and $k_{\rm{spring}}$ and their effect on CNCC migration in the model. A) When $f_0$ is low, the active force induced by VEGF gradients is insufficient to drive migration along the domain. B) When both $f_0$ and $k_{\rm{spring}}$ are sufficiently large, the force induced by VEGF gradients is sufficient to drive migration along the domain, leading to typical CNCC migratory formations. C) If $f_0$ is much larger than $k_{\rm{spring}}$, the active force induced by VEGF gradients leads to the escape of highly polarised cells at the leading edge of collectives, resulting in aberrant migration. Adjacent to each schematic are kymographs displaying the density of cells as a function of $x$ in numerical model simulations when migration ends at $t=24$h (500 simulation repeats).}
    \label{fig:Figure3}
\end{figure}

For intermediate values of $\lambda$ that facilitate persistent polarisation at the leading edge of streams, the model exhibits three distinct migratory behaviours determined by the interplay between the magnitude of intercellular adhesion, $k_{\mathrm{spring}}$, and the active chemotactic force scale, $f_0$. Too low a value of $f_0$ means that invasion is inhibited irrespective of the value of $k_{\mathrm{spring}}$ as there is an insufficient chemotactic drive at the leading edge of streams to propel collectives forward (Figure \ref{fig:Figure3}A). As $f_0$ is increased, two distinct phases of migration emerge. Too low a value of $k_{\mathrm{spring}}$ relative to $f_0$ leads to the escape of a small subset of cells at the leading edge, when strong chemotactic polarisation induces a high active force away from the weakly-polarised and loosely-connected bulk (Figure \ref{fig:Figure3}C).  
However, for sufficiently high $f_0$ and $k_{\mathrm{spring}}$, the active force generated by highly polarised cells is sufficient to propel collectives along the domain without compromising intercellular adhesions that are essential to maintain a cohesive stream (Figure \ref{fig:Figure3}B). 

Thus, model simulations indicate that balanced levels of active chemotactic force generation and intercellular adhesion are required for sustained, coherent migration as a collective. Similar trends have been observed both experimentally in the chick embryo and in a corresponding ABM of CNCC migration \cite{mclennan2020neural}. \textit{In vivo}, gain-of-function of Aquaporin-1, a gene linked to filopodial stabilisation and persistent migration, increases the distance of CNCC invasion, while a loss-of-function results in failed invasion of ba2. In a corresponding ABM, these perturbations manifest as leader cell escape under gain-of-function and failed invasion under loss-of-function, respectively.

\subsection{Leader and follower phenotypes emerge spontaneously in a spring-polarisation ABM of CNCC migration}
\label{emergence}
To further investigate the behaviour of the model formulated here, we select a combination of $\{f_0, k_{\mathrm{spring}}, \lambda\}$ that results in invasion over biologically relevant distances and that produces coherent CNCC streams which remain connected throughout migration. 
A model parameter sweep (Figure~\ref{fig:sweep}B) reveals that a range of $\{f_0, k_{\mathrm{spring}}, \lambda\}$ triplets yield the desired outcome, in which collective migration is sustained over a period of 24h, and the maximum invasion distance of the stream is comparable to observations \textit{in vivo} ($x_{\rm{max}}>800\mu\rm{m})$. As such, we select the parameter values $\{f_0=1000\mu\rm{m\,h^{-1}}, k_{\mathrm{spring}}=100\rm{h^{-1}}, \lambda=300\rm{h^{-1}}\}$, which lie well within these intervals, to serve as a representative regime for further analysis. This regime supports sustained, cohesive migration over biologically relevant timescales and distances, enabling further investigation of the model and its emergent behaviour.

\begin{figure}
    \centering
    \includegraphics[width=\linewidth]{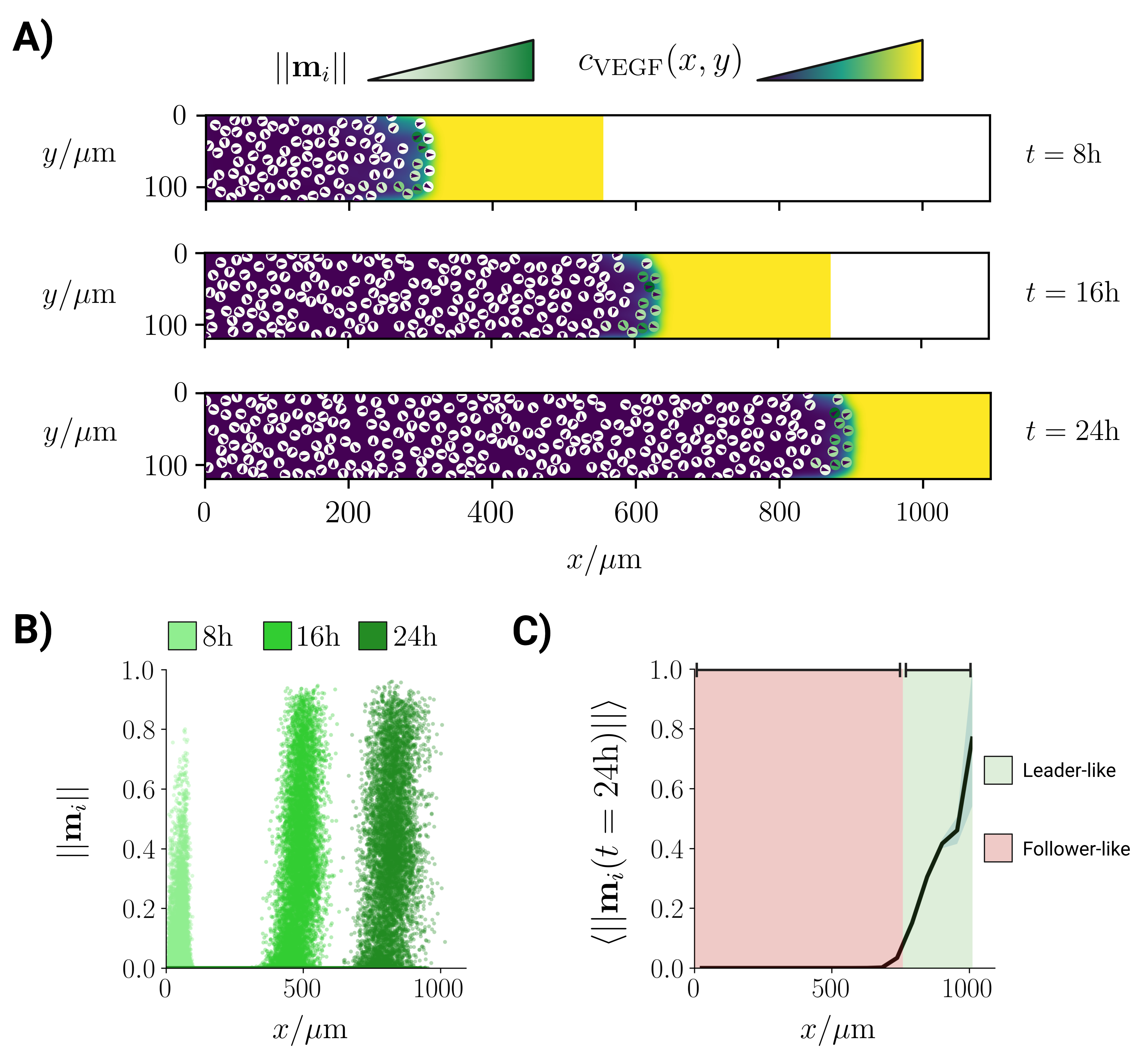}
    \caption{A) Typical model simulation for $\{f_0=1000\mu\rm{m\,h^{-1}}, k_{\mathrm{spring}}=100\rm{h^{-1}}, \lambda=300\rm{h}^{-1}\}$. Initially, CNCCs emerge into regions of high VEGF concentration, and thus, begin to generate large chemical gradients as VEGF is degraded. The generation of large VEGF gradients induces a strong chemical polarisation in CNCCs, which results in their movement along the domain. Cells emerging at later times emerge into regions of depleted VEGF, and therefore, remain largely unpolarised. As such, their movement is primarily governed by intercellular attraction-repulsion mechanisms. Throughout migration, CNCCs at the front of the collective remain highly polarised, thus creating a persistent subset of cells for which chemotaxis is the primary determinant of directional movement. B) Scatter plot of $||\mathbf{m}_{i}||$ for $\{f_0=1000\mu\rm{m\,h^{-1}}, k_{\mathrm{spring}}=100\rm{h^{-1}}, \lambda=300\rm{h}^{-1}\}$ at $t=8\rm{h}, 16\rm{h}$, and $24\rm{h}$. Cells at the leading edge of streams are highly polarised due to consistent exposure to large VEGF gradients. Cells behind the leading edge remain unpolarised. C) $\left<||\mathbf{m}_{i}(t=24\rm{h})||\right>$ as a function of $x$ for the same data set as B). Non-zero mean polarisations are restricted to the green region at the front of the stream. All other cells in the red region are largely unpolarised when migration ends. The data in Figures B) and C) were collected over 500 model simulations.}
    \label{fig:Figure4}
\end{figure}

Studying the magnitude of the cell-specific polarity vector, $||\mathbf{m}_{i}||$, as a function of the distance migrated away from the embryonic hindbrain at $x=0$, reveals the emergence of a phenotypic division between cells at the front of streams and trailing cells behind. 
As migrating CNCCs locally degrade VEGF, cells at the leading edge consistently invade regions where VEGF gradients are large (Figure \ref{fig:Figure4}A).
These gradients drive stronger chemotactic activation in cells at the leading edge, resulting in higher chemical-induced polarity and larger active forces that propel the movement of collectives.
In contrast, trailing cells migrate through regions of depleted VEGF, and therefore, remain weakly-polarised, moving primarily through mechanical interactions with their neighbours.
This effect results in a subset of cells at the leading edge of streams \corr{becoming polarised upon entry into the domain} and remaining polarised ($||\mathbf{m}_{i}|| > 0$) thereafter, while all other cells remain largely unpolarised, ($||\mathbf{m}_{i}|| \approx 0$) (Figure \ref{fig:Figure4}B). 

Interpreting these results biologically suggests that leader-like and follower-like phenotypes emerge spontaneously during the collective migration of CNCCs. Cells that remain strongly polarised during migration move primarily according to chemical-induced motility, mirroring leader-like cells in the chick embryo for which transcriptomic analysis suggests an increased tendency to move according to gradients in VEGF \cite{mclennan2012multiscale}. Cells that remain largely unpolarised behind the leading edge of streams receive movement cues primarily from neighbouring cells, reflective of follower-like cells in the chick embryo that up-regulate genes associated with cell--cell adhesion (e.g. Cadherins \cite{mclennan2012multiscale}). Defining cells with $||\mathbf{m}_{i}||>10^{-1}$ as leader-like cells, and all other cells as follower-like (Figure \ref{fig:Figure4}C), model simulations demonstrate the emergence of strong spatial correlation in phenotypes, which reflects observations from the chick embryo. These findings lend credibility to existing models of leader--follower migration, but demonstrate the existence of a narrow region of continuous, rather than binary, phenotypic variation at the leading edge of streams, and that these phenotypic differences may emerge spontaneously from both environment and intercellular guidance cues that are detected throughout migration. 

\subsection{A spring-polarisation ABM of CNCC migration produces migratory behaviour consistent with experimental perturbations in the chick embryo}
\label{experiments}

\begin{figure}
    \centering
    \includegraphics[width=\linewidth]{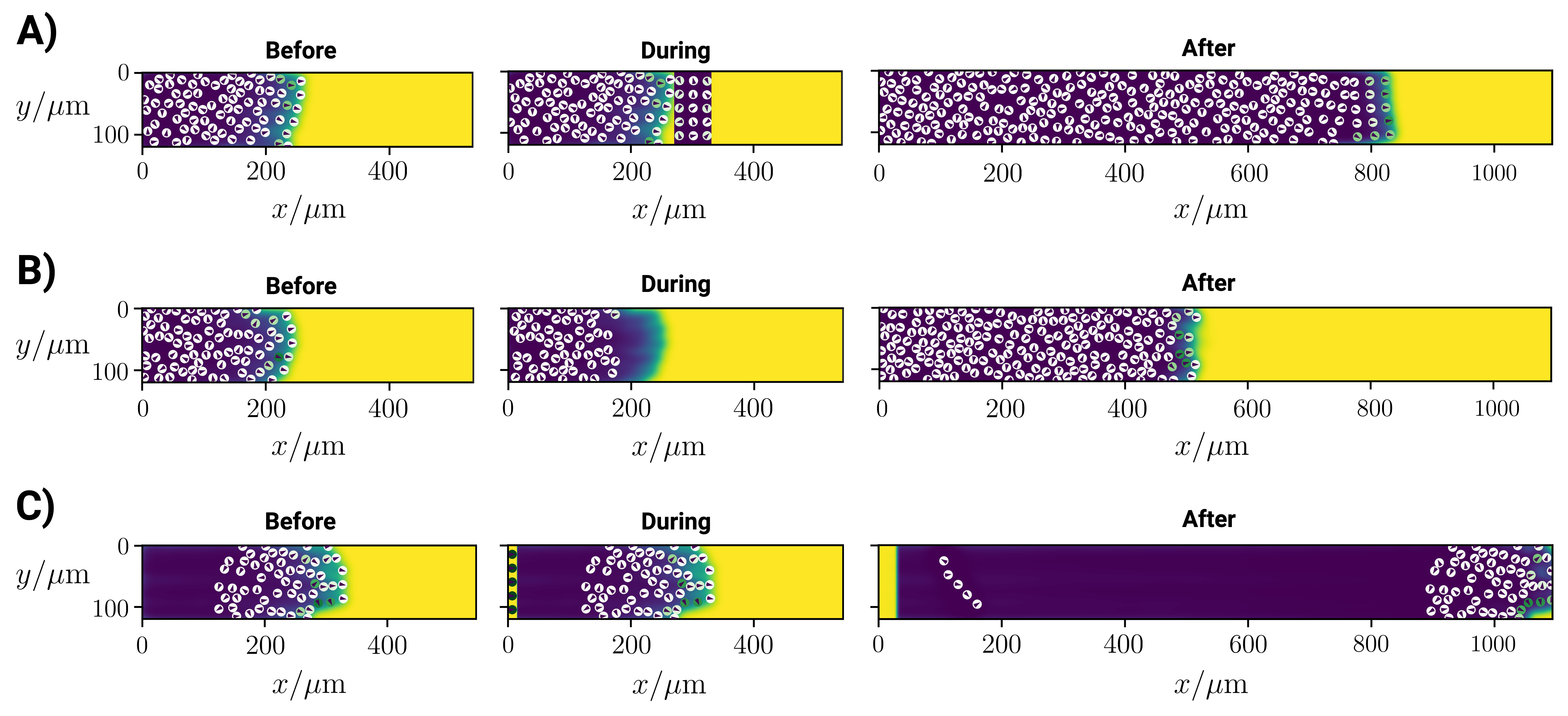}
    \caption{The results of perturbations to numerical model simulations based on prior experiments conducted in the chick embryo \cite{mclennan2012multiscale, richardson2016leader}. A) Upon the transplantation of tissue containing $n=15$ unpolarised cells ahead of the leading edge of CNCC streams $12$h after migration begins, migration transiently ceases, as there are no longer polarised cells at the leading edge to guide migration. After a short transient period, cell-induced VEGF gradients form in the region of transplantation. Subsequently, transplanted cells become polarised, such that migration resumes in a typical stream formation. B) Upon the ablation of $n=20$ cells at the leading edge of CNCC streams $12$h after migration begins, the remaining cells at the leading edge of the post-ablation stream are exposed to local gradients in VEGF, which drives their polarisation and a subsequent resumption in active migration. C) When polarised cells are transplanted into the region of the r4--ba2 pathway adjacent to the embryonic hindbrain $12$h after migration begins, they are positioned in a region of degraded VEGF and, therefore, cannot gain the polarisation required to invade ba2.} 
    \label{fig:Figure5}
\end{figure}

Ablation and transplantation experiments conducted in the chick embryo \cite{mclennan2012multiscale, richardson2016leader} have yielded a range of testable predictions regarding the behaviour of CNCC streams in response to environmental and cellular perturbations. To test the validity of the model formulated here, we now compare numerical model simulations with the outcomes of these experiments. The results of each experiment are visualised in Figure \ref{fig:Figure5}.  As before, we take representative values of the parameters governing cell--cell and chemotactic forces $\{f_0=1000\mu\rm{m\,h^{-1}}, k_{\mathrm{spring}}=100\rm{h^{-1}}, \lambda=300\rm{h}^{-1}\}$ with other parameter values as described in Appendix~B. 

\subsubsection{Collective CNCC migration persists upon the transplantation of trailing cells to the leading edge of streams}
In experiments where unpolarised follower-like cells are transplanted at the leading edge of streams, it has been observed that collective migration stalls for a short transient period, before further migration towards ba2 \cite{mclennan2012multiscale}.  

In model simulations where $n=15$ unpolarised cells are transplanted ahead of the leading edge of CNCC streams $12$h after migration begins, a similar phenomenon is observed (Figure \ref{fig:Figure5}A). Initially, the unpolarised cells cannot generate sufficient active forces to propel the collective towards ba2. However, as VEGF is degraded in the region of transplantation, transplanted cells begin to gain polarisation, which induces migration towards ba2, as observed \textit{in vivo}. 

In a previous ABM of CNCC migration, this phenomenon could not be fully captured \cite{mclennan2012multiscale}, with further modifications to the model required to recapitulate the behaviour observed \textit{in vivo}. In the spring--polarisation model studied here, this behaviour is robust to both the number of cells transplanted and the time of transplantation, provided that the leading edge of the stream after transplantation lies in regions where VEGF has not been degraded. Biologically, these results support a self-organised model of CNCC streams, in which directed collective movement arises from local, context-dependent interactions rather than from the fixed leader--follower hierarchy considered in prior ABMs of CNCC migration  \cite{mclennan2012multiscale, mclennan2015vegf, mclennan2023colec12, mclennan2020neural, mclennan2017dan, johnson2025mathematical}.

\subsubsection{Collective CNCC migration persists upon the ablation of cells at the leading edge of streams}
In the chick embryo, collective CNCC migration persists upon the ablation of CNCCs at the leading edge of streams \cite{richardson2016leader}. This observation was used to argue that leader-like CNCCs do not constitute a fixed sub-population and that all CNCCs within a stream share an equivalent capacity to adopt leader-like behaviours. Here, we test this hypothesis by ablating $n=20$ cells at the leading edge of the stream $12$h after migration begins.

Numerical simulations of this experiment align with observations in the chick embryo (Figure \ref{fig:Figure5}B). When the twenty front-most cells in CNCC streams are ablated, the cells at the leading edge of the post-ablation stream lie in regions of detectable VEGF gradients. As such, after a transient period without active migration, these cells become polarised and once again propel the collective towards ba2. 

However, varying the number of cells ablated at the leading-edge reveals that this behaviour does not persist when larger numbers of cells are ablated. For migration to persist upon leading cell ablation, cells at the leading edge of the ablated stream must remain within regions where a detectable VEGF gradient is present, such that they can become polarised. This is true for cells immediately behind the original leading edge of the stream, but not in positions closer to the embryonic hindbrain, where VEGF has largely been degraded. Consequently, extensive frontal cell ablation shifts the leading edge of the stream into a region void of chemotactic cues, resulting in a persistently unpolarised bulk that cannot reinitiate movement. Biologically, these findings indicate that, although all CNCCs possess the capacity to acquire leader-like properties, this transition is not determined solely by intrinsic cellular properties, but rather by cell–environment interactions. The results of varying the number of leading cells ablated at $t=12\rm{h}$ are shown in Appendix~C. 

\subsubsection{Cells at the leading edge of streams transplanted to the region adjacent to the embryonic hindbrain fail to invade ba2}

In experiments, leader-like cells transplanted into the region adjacent to the embryonic hindbrain fail to invade ba2, and instead remain in place \cite{mclennan2012multiscale}. This phenomenon is also observed in numerical model simulations, provided that the transplanted cells lie outside of the range of cell--cell interaction of the stream (Figure \ref{fig:Figure5}C). 

When highly polarised cells are transplanted behind the stream, they lie in regions where VEGF has been eroded, and thus, begin to lose polarisation after a transient period of migration. As they are sufficiently far from the stream, they also cannot receive cell--cell guidance cues from cells within the main stream. As such, they remain in place, and are unpolarised when migration ends. 

Taken together, numerical simulations of experiments conducted in the chick embryo show concurrence with behaviour \textit{in vivo}. As such, the model formulated here provides a unified mechanistic mathematical framework that is consistent with both experiments and unperturbed migration, suggesting that it is an appropriate model framework within which to study collective migration. 

\section{Conclusions}

We have introduced a polarity-based ABM for CNCC migration that captures the spontaneous emergence of leader–follower behaviour without imposing phenotypic differences \textit{a priori}. In this framework, all cells evolve according to identical rules, interact mechanically through spring-like couplings, and respond to time-averaged self-induced gradients in VEGF according to an ODE for chemical-induced polarisation. Numerical simulations of the model show that, given these mechanisms, distinct leader-like and follower-like behaviours emerge spontaneously from the collective dynamics of the system \corr{according to differences in chemical-induced polarity}. These findings demonstrate that persistent heterogeneity in migratory behaviour can arise solely from the interplay of chemotactic sensing, persistence in polarisation, and mechanical cell--cell couplings, without requiring pre-determined cell identities at the point of delamination from the embryonic hindbrain.

The model reproduces a range of experimental observations from \textit{in vivo} and \textit{in vitro} studies of CNCC migration. Leader-like cells at the front of streams align with local VEGF gradients and direct collective movement, while follower-like cells behind them maintain stream cohesion through mechanical coupling and contact guidance. Stream-like structures remain coherent over biologically relevant timescales and respond adaptively to heterogeneous chemical landscapes. As such, directed collective migration is achieved without imposing an intrinsic spatial bias or anisotropic motility parameters, \corr{which has been a necessary assumption in previous ABMs for them to reproduce the stream-like migratory patterns observed \textit{in vivo}}. This, in turn, suggests that emergent polarity dynamics and isotropic cell--cell interactions alone are sufficient to sustain coordinated migration in the cranial neural crest.

From a modelling perspective, this work extends previous ABMs of CNCC migration by relaxing both the assumption of a leader--follower phenotypic dichotomy and also of instantaneous responses to environmental perturbations made in a later signal integration ABM of CNCC migration \cite{schumacher2019neural}. Incorporating persistence in polarity provides a mechanistic link between finite-interval environmental sensing and cell migration, allowing the model to capture both transient environmental adaptation and long-term behavioural stability within a unified framework.

Future work could extend the present model in several directions. Coupling polarity dynamics to intracellular signalling networks may help connect molecular-scale regulation to emergent migratory behaviours \cite{welf2011signaling}. Additionally, incorporating interactions with the ECM or heterogeneous mechanical environments could further improve biological realism, as cell--ECM interactions are a well-documented mechanism of migration for CNCCs \textit{in vivo} \cite{bilozur1988neural, perris1989molecular}. Finally, the framework developed here could be generalised to study other instances of collective migration where leader--follower organisation emerges spontaneously, such as cancer metastasis \cite{deisboeck2009collective} or wound healing \cite{li2013collective}.

\subsubsection*{In memoriam}
This research was carried out shortly after the passing of Paul Kulesa (1962--2025). Paul was a long-standing collaborator of the Wolfson Centre for Mathematical Biology (WCMB), and his pioneering work on cranial neural crest biology laid the foundations for many subsequent studies in which experimental and mathematical perspectives have informed one another to deepen our understanding of collective cell migration. Without his insights and scientific leadership, this research, and much of the work it builds upon, would not have been possible. Paul’s influence on the field remains profound, and he lives on in the research he continues to inspire. He is deeply missed by all of us in the WCMB who had the privilege of working with him. Paul is survived by his wife, Jennifer, and their five children Eliason, Wesley, Wallace, Watson, and Roxy.

\subsubsection*{Conflict of interest statement} 
The authors declare no conflict of interest.

\subsubsection*{Authors' contributions}
\textbf{Samuel Johnson:} Conceptualisation (lead), data curation (lead); formal analysis (lead); methodology (lead); software (lead); writing–original draft (lead); writing–review and editing (supporting). \textbf{Ruth E. Baker}:  Conceptualisation (supporting), supervision (lead);  writing–review and editing (lead). \textbf{Philip K. Maini:}  Conceptualisation (supporting), supervision (lead);  writing–review and editing (lead).

\subsubsection*{Acknowledgments}
S.W.S.J. receives support from the Biotechnology and Biological Sciences Research Council (BBSRC) (grant number BB/T008784/1). R.E.B. is supported by a grant from the Simons Foundation (MP-SIP-00001828). For the purpose of open access, the authors have applied a CC BY public copyright licence to any author accepted manuscript arising from this submission.

\bibliography{BIBL}
\bibliographystyle{unsrt}

\end{document}


\appendix

\renewcommand{\thefigure}{S\arabic{figure}}

\title{Appendix: The spontaneous emergence of leaders and followers in a mathematical model of cranial neural crest cell migration}
\author[1]{Samuel W.S. Johnson\thanks{Corresponding author: samuel.johnson@chch.ox.ac.uk}}
\newcommand\CoAuthorMark{\footnotemark[\arabic{footnote}]}
\author[1]{Ruth E. Baker\thanks{These authors contributed equally.}}
\author[1]{Philip K. Maini\protect\CoAuthorMark}
\affil[1]{Wolfson Centre for Mathematical Biology, Mathematical Institute, University of Oxford, Oxford, United Kingdom}
\date{}
\maketitle

\linespread{1.5}\selectfont

\section{Model implementation}
We outline implementation details of the spring--polarisation model, before listing the values of fixed model parameters in numerical simulations. 

\subsection{Domain growth}
The r4--ba2 pathway through which CNCCs migrate is represented by a two-dimensional rectangle of a fixed width ($L_{y} = 120\mu$m) and a time-dependent length, $L_{x}(t)$. The domain grows logistically in time and in a spatially uniform manner. Our implementation of domain growth is based on data collected from the chick embryo \cite{mclennan2012multiscale}. The length of the domain as a function of time is given by

\begin{equation}
    \label{domainGrowthEquation}
    L_{x}(t) = \frac{L_{0}}{1 + e^{-k(t - t_0)}} + b,  
\end{equation}
\\
\noindent where $L_0 = 855.8 \mu $m, $k = 0.294 \text{h}^{-1}$, $t_0=14.98$h, and $b=294.3\mu $m. Parameter values were obtained by fitting the experimental data to Equation \eqref{domainGrowthEquation}.

\subsection{Partial differential equation for the time evolution of VEGF}

In Section 2.1.2, Equation~(1) gives the non-dimensional, re-scaled form of the governing equation for VEGF dynamics. Here, we present the corresponding dimensional form prior to re-scaling. The concentration of VEGF, $c(x,y,t)$, evolves according to

\begin{equation}
\label{chemicalEquation}
\frac{\partial c}{\partial t}
+
\underbrace{\frac{\partial (a c)}{\partial x}}_{\text{(I)}}
=
\underbrace{D\left(\frac{\partial^2 c}{\partial x^2} + \frac{\partial^2 c}{\partial y^2}\right)}_{\text{(II)}}
-
\underbrace{c \sum_{i=1}^{N(t)} \frac{\lambda}{2\pi R^2}
\exp\!\left[-\frac{(x-x_i)^2 + (y-y_i)^2}{2R^2}\right]}_{\text{(III)}}
+
\underbrace{\kappa c(1-c)}_{\text{(IV)}},
\end{equation}

\noindent where term (I) represents advection and dilution of VEGF associated with elongation of the tissue domain at rate $a=a(x,t)$, term (II) represents diffusion with diffusion coefficient $D$, and term (III) models VEGF internalisation and degradation by cranial neural crest cells (CNCCs);  
each cell, indexed by $i$ and at a position $(x_i, y_i)$, locally depletes VEGF according to a Gaussian kernel of width $R$ (representing the cell radius) and a maximum rate of degradation, $\lambda>0$. Term (IV) represents spatially uniform VEGF production at a rate $\kappa$, with a logistic growth law to constrain the concentration to $0 \leq c(x,y,t) \leq 1$. \\

\noindent Zero-Dirichlet boundary conditions are imposed at $x = 0$ and $x = L_x(t)$ to avoid artificial polarisation of cells towards the domain boundaries upon entry, while periodic boundary conditions are imposed at $y=0$ and $y=L_{y}$ to represent the presence of adjacent tissue, into which VEGF may diffuse.

\subsection{VEGF gradient sensing accuracy}
Our implementation of VEGF gradient detection is based on work by Berg and Purcell \cite{berg1977physics} and is as described in previous ABMs of CNCC migration (e.g. \cite{mclennan2012multiscale}). In the model, a filopodial protrusion extended at an angle $\phi$ senses a chemical gradient when 

\begin{equation}
\frac{c_{\rm f}(\phi_{\rm{max}})-c(\mathbf{x}_i)}{c(\mathbf{x}_i)}
> 
\xi\,\sqrt{\frac{c_0}{c(\mathbf{x}_i)}},
\label{chemoreception}
\end{equation}
\\
\noindent where $c_{\rm f}(\phi)$ is the average VEGF concentration sensed by a filopodium extended at an angle $\phi$, $c(\mathbf{x}_i)$ is the VEGF concentration at cell $i'$s current position, $\xi$ is a dimensionless parameter governing gradient detection, and $c_0$ is the initial uniform VEGF concentration in non-dimensional units. 
\\

\noindent An issue with this implementation in numerical model simulations is that both sides of Equation~\eqref{chemoreception} diverge as $c(\mathbf{x}_i)\to 0$, which leads to unrealistically high chemotactic sensitivity at very low VEGF concentrations. To prevent this phenomenon in simulations, we impose a lower bound on the effective concentration used in the detection criterion and modify the condition for gradient detection to 
\\
\begin{equation}
\frac{c_{\rm f}(\phi_{\rm{max}})-c(\mathbf{x}_i)}{\rm{max}\{c(\mathbf{x}_i), \epsilon\}}
> 
\xi\,\sqrt{\frac{c_0}{\rm{max}\{c(\mathbf{x}_i), \epsilon\}}},
\end{equation}
\\
\noindent where $\epsilon=10^{-2}$ is an effective lower bound on the sensed concentration that prevents the detection of small gradients in low VEGF concentrations.

\section{Model parameterisation}
Parameter values in the PDE governing VEGF dynamics and the properties of cells are primarily derived from prior ABMs for CNCC migration. Table \ref{parameters} lists model parameters that are fixed in numerical model simulations, with Section \ref{notes} providing justification for parameter choices made in the absence of prior mathematical or biological parameterisation.

\vspace{12pt}

\begin{table}[ht!]
\small
\begin{tabular}[t]{p{1.1cm} p{6cm} p{3cm} p{5cm}} 
& \textbf{Description} & \textbf{Value} & \textbf{Reference} \\
\hline$R$ & cell radius (nuclear) & $7.5 \mu \mathrm{m}$ & McLennan \& Kulesa $(2010)$ \cite{mclennan2010vascular} \\
$L_{y}$ & height of domain & $120 \mu \mathrm{m}$ & McLennan et al. (2012) \cite{mclennan2012multiscale} \\
$L_{x}$ & length of domain (Eq. (A)) & $294\mu$m to $1107 \mu$m & McLennan et al. $(2012)$ \cite{mclennan2012multiscale}\\
$\ell_{\rm f}$ & cell sensing radius & $27.5 \mu \mathrm{m}$ & McLennan et al. $(2017)$ \cite{mclennan2017dan} \\
$\ell_{\rm f}^{\rm{max}}$ & maximum distance for intercellular bonding & $45 \mu \mathrm{m}$ & McLennan et al. $(2017)$ \cite{mclennan2017dan} \\
$D$ & VEGF diffusivity & $0.1 \mu \mathrm{m}^{2} / \mathrm{h}$ & McLennan et al. $(2017)$ \cite{mclennan2017dan}\\
$\kappa$ & VEGF production rate & $10^{-4} / \mathrm{h}$ & McLennan et al. $(2017)$ \cite{mclennan2017dan}\\
$\xi$ & chemical sensing accuracy & 0.25 & McLennan et al. $(2017)$ \cite{mclennan2017dan} \\
$\tau_{\text{sense}}$
& relaxation time for polarity vector 
& $0.10\mathrm{h}$ 
& McLennan et al. (2015) \cite{mclennan2015vegf} \\
$k_{\text{chemo}}$ & angular relaxation rate & $10\mathrm{h}^{-1}$ 
& McLennan et al. (2015) \cite{mclennan2015vegf} \\
$n_{\text {f}}$ & number of directions sampled in one sampling interval & 5 & N/A -- Section B.1 \\
$k_{\text{rep}}$ & strength of short-range repulsive interaction  & $10^{3}\rm{h}^{-1}$ &  N/A -- Section B.1 \\
$\gamma_{\rm{sat}}$ & chemotactic detection saturation parameter & 0.5 & N/A -- Section B.1 \\
$\eta$ & stochastic positional noise amplitude & 0.1 & N/A -- Section B.1 \\

\end{tabular}
\caption{Fixed parameter values in the ABM, along with references, where appropriate.}
\label{parameters}
\end{table}
\normalsize

\subsection{Notes on model parameterisation}
\label{notes}
We provide brief notes on fixed model parameters that cannot be derived from the existing biological and mathematical modelling literature. 

\subsubsection{Number of directions sampled in one sampling interval}
\corr{\textit{In vivo}, CNCCs project many filopodia along their membranes \cite{genuth2018chick}. These filopodia typically persist for time intervals that are longer than the interval between successive environmental sampling events in the model. Accordingly, we set $n_{\rm f} = 5$ in the model, which is selected to balance the smaller number of filopodia extended by CNCCs in the model against the higher rate of sampling compared with CNCCs \textit{in vivo}. This parameter value is similar to previous ABMs of CNCC migration \cite{mclennan2012multiscale, mclennan2015vegf, mclennan2023colec12, mclennan2020neural, mclennan2017dan, johnson2025mathematical, schumacher2019neural}.}

\subsubsection{Strength of short-range repulsive interactions}
There is no way to directly parameterise $k_{\rm{rep}}$, the Hookean spring constant governing the force between overlapping cells. As such, we fix $k_{\rm{rep}} = 10^{3}\rm{h}^{-1}$ in model simulations, which is sufficiently large to prevent cells overlapping during migration. However, all model predictions are robust to large variations in this parameter. 

\subsubsection{Chemotactic detection saturation parameter}
The parameter $\gamma_{\rm{sat}}$ governs how rapidly the intracellular signals induced by VEGF gradients saturate at a maximum. There is no way to parameterise this biologically, but we find that the qualitative trends observed in numerical model simulations are robust to variations in this parameter ($\gamma_{\rm{sat}} \in [ 0.25, 0.75]$). 

\subsubsection{Stochastic positional noise}

The parameter $\eta$ sets the amplitude of the stochastic positional perturbations applied during each numerical substep of the overdamped mass--spring update. In the implementation, a two-dimensional random vector with independently and uniformly sampled components in $[-1,1]$ is generated at every time step. This vector is then multiplied by the scalar amplitude, $\eta$, before being added to the cell's current position. This stochastic contribution represents unresolved microscopic variability in the microenvironment through which migration occurs. In simulations of the model, we fix $\eta = 0.1$. However, model predictions remain consistent provided the noise amplitude remains small compared to deterministic displacements arising from chemotactic polarisation and cell--cell mechanical interactions.

\subsubsection{Adhesive bond formation and removal}

Cell--cell adhesion is implemented via transient adhesive bonds that form and break dynamically during migration. When the distance between two cells falls below the interaction radius $\ell^{\max}_{\rm f}$, an adhesive bond is formed and represented as a Hookean spring with a rest length equal to the inter-cell separation at the time of bond formation. The choice of $\ell^{\max}_{\rm f}$ is informed by prior agent-based models of CNCC migration \cite{johnson2025mathematical, mclennan2017dan}. \\

\noindent Adhesive bonds are removed once the inter-cell distance exceeds a prescribed breaking radius, taken to be slightly larger than the interaction radius ($1.1\,\ell^{\max}_{\rm f}$). Introducing a distinct breaking threshold reduces spurious bond turnover for cell pairs fluctuating near the maximum range of cell--cell interaction and ensures stable adhesive coupling over biologically relevant timescales.

\section{Varying the number of leading cells ablated at $t=12\rm{h}$}

\begin{figure}[h!]
    \centering
    \includegraphics[width=0.5\linewidth]{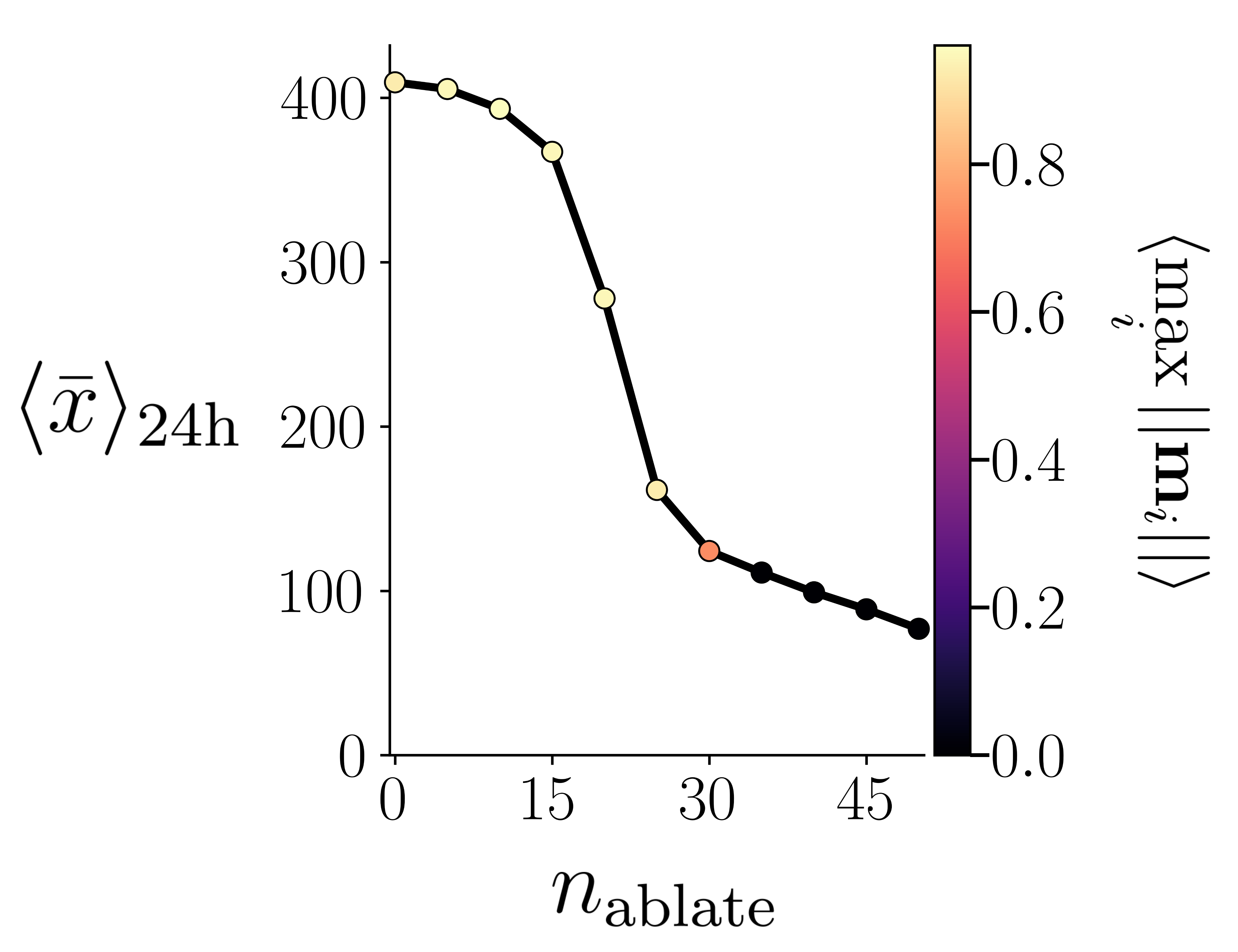}
    \caption{Mean stream invasion distance (average $x$-coordinate of stream centroids) as a function of the number of leading cells ablated. Points show the mean over 500 simulations and point colour denotes the mean maximum polarisation magnitude within streams. When only a small number of cells are ablated, migration persists over biologically relevant distances, giving an $x$-centroid comparable to unperturbed migration. As the ablation size increases, cells at the leading edge lie in regions of eroded VEGF, fail to polarise, and invasion is reduced.}
    \label{fig:ablationSweep}
\end{figure}

Ablating cells at the leading edge of streams at $t=12\rm{h}$ reveals that the subsequent migratory behaviour of streams varies with the number of leading cells ablated. For small numbers of ablated cells, cells at the leading edge of the ablated stream lie in regions of large VEGF gradients, such that they subsequently become polarised and reinitiate migration (Figure \ref{fig:ablationSweep}). However, the ablation of a larger number of cells pushes the leading edge into regions where VEGF has largely been eroded. As such, all cells remain unpolarised and invasion does not occur over biologically relevant distances. 

\newpage 

\bibliography{BIBL}
\bibliographystyle{unsrt}